# Beyond Electrons: A Theoretical Framework for Near-Field Radiative Thermal Computing and Neural-Network-Inspired Processing

**Hexiang Zhang[1], Mauro Antezza[3,4] and Yi Zheng[1,2,*]**

[1] Department of Mechanical and Industrial Engineering, Northeastern University, Boston, MA 02115, USA

[2] Department of Chemical Engineering, Northeastern University, Boston, MA 02115, USA

[3] Laboratoire Charles Coulomb (L2C), UMR 5221 CNRS-Université de Montpellier, F-34095 Montpellier, France

[4] Institut Universitaire de France, 1 rue Descartes, Paris Cedex 05, F-75231, France

** Correspondence: Yi Zheng*

## Abstract

Near-field radiative heat transfer provides a route for information processing in which thermal radiation, rather than charge transport, serves as the physical carrier of signals. Here, we propose and theoretically analyze a programmable near-field radiative thermal computing framework in which radiative coupling, phase-change nonlinearity, and thermal-state memory are mapped onto neural-network-inspired operations. The framework is constructed from near-field radiative thermal diodes, transistors, and multi-terminal logic units separated by nanoscale gaps. Radiative heat flux represents the propagated thermal information, while geometry- and material-dependent radiative coupling provides physically constrained weighting, and the temperature-dependent optical response of phase-change materials enables nonlinear modulation and logic-state control. Based on these primitives, we formulate a radiative thermal convolutional network for spatial information processing and a radiative thermal recurrent network for history-dependent computation. The recurrent response is associated with radiative feedback, thermal relaxation, and phase-change hysteresis, with $VO_2$ providing history-dependent short-term memory and GST offering a possible route toward non-volatile phase storage. We further distinguish the physical radiative networks from a separate software-based inverse-identification study, in which recurrent machine-learning models are trained on simulated near-field heat-flux–temperature characteristics to recover structural parameters. The present work is therefore intended as a theoretical and numerical framework rather than an experimental realization. By establishing a bottom-up connection between fluctuational electrodynamics, radiative thermal logic, programmable thermal states, and neural-network-inspired computation, this study provides a physically grounded basis for exploring non-contact thermal information processing at the near-field limit.

## Introduction

Artificial neural networks are conventionally realized through electronic hardware in which charge, voltage, and current represent information and programmable electrical conductances represent connection weights. The extraordinary success of electronic computing does not imply that information processing must be confined to charge transport. A broader physical-computing perspective asks whether other conserved or transported quantities can encode, transform, and retain information directly through their governing physics. Heat is particularly interesting in this context because temperature fields and thermal fluxes are already the native observables in thermal sensing, infrared detection, energy systems, and many extreme-environment applications.

Thermal logic is not itself a new concept. Conductive thermal diodes, thermal transistors, phase-change switches, thermal memories, and thermal-electric neuromorphic elements have established that nonlinear heat transport can support switching and state-dependent operations. However, this contribution considered here is narrower and more specific: we ask whether near-field radiative heat transfer can provide a programmable, non-contact coupling mechanism from which spatial weighting, nonlinear processing, logic decoding, and recurrent thermal memory can be constructed in a common physical framework. [6,7,10,13,27,38-40,44] When two bodies are separated by a nanoscale vacuum gap, evanescent electromagnetic modes can tunnel across the separation and produce radiative heat transfer far above the far-field blackbody limit. More importantly for information processing, the magnitude and spectrum of this transfer depend strongly on gap distance, dielectric response, surface and bulk polaritonic resonances, geometry, and material phase. Near-field radiative transfer therefore offers more than a large heat flux: it provides a state-dependent electromagnetic coupling that can be altered by temperature, phase-change state, patterning, and geometry. [16,20,22-24,26,31,47]

In the present architecture, localized source, gate, and drain thermal terminals are separated by nanoscale gaps. Polar dielectric layers such as h-BN support strong near-field electromagnetic coupling, while phase-change materials such as $VO_2$ or GST modify the optical response of a gate terminal. The resulting heat flux is treated as the propagated thermal signal. A local differential response of heat flux to temperature defines a physically realizable coupling coefficient. This coefficient is not an arbitrary software weight; it is constrained by fluctuational electrodynamics, material properties, geometry, passivity, energy conservation, and, where applicable, reciprocity. [17,41-43] On this basis, we construct two network-level concepts. The first is a physical radiative thermal convolutional network, in which a finite local receptive field and a spatially repeated radiative coupling pattern provide the basis of a convolution-like operator. The second is a physical radiative thermal recurrent network, in which finite thermal capacitance, radiative feedback, and PCM state evolution cause the present thermal response to depend on previous thermal states. The recurrent formulation is derived from thermal energy balance rather than imposed as a direct copy of a software recurrent equation.

A central distinction of the framework is the separation between physical forward computation and numerical optimization. The physical network can perform a programmed radiative forward operation once its gate temperatures, phase states, and geometrical parameters have been set. Gradient-based learning, however, is treated as an offline numerical method for finding physically admissible programming parameters unless an independent in-situ learning mechanism is explicitly demonstrated. Likewise, a separate software LSTM/GRU inverse-identification study is treated as a machine-learning application to simulated radiative data rather than as proof that the physical thermal network implements LSTM or GRU gates. This reframing also clarifies the scope of the study. The work is theoretical and numerical. It does not report a fabricated integrated radiative neural processor, autonomous physical backpropagation, or a general replacement for electronic computing. Instead, it establishes a bottom-up physical mapping from fluctuational electrodynamics to device-level thermal logic, local spatial processing, temporal recurrence, and memory. Such a framework may be especially relevant when information already exists in the thermal or infrared domain, where local processing prior to full thermal-to-electrical conversion could provide a specialized advantage. The rest of the manuscript develops this connection step by step. We first establish the physical basis of near-field radiative transport and define the thermal information variables. We then formulate radiative thermal diodes, gate-controlled modulators, and multi-terminal logic units as computing primitives. These devices are mapped to physically constrained weighted and nonlinear operations, followed by spatial T-CNN and temporal T-RNN architectures, optimization rules, proof-of-principle numerical tests, a separate software inverse-identification task, and finally a discussion of robustness, scalability, and practical implementation.

## Physical Basis of Near-Field Radiative Thermal Computing

Near-field radiative thermal computation is governed by the same statistical and electromagnetic principles as near-field radiative heat transfer. The computational interpretation begins only after the heat-transfer quantities have been defined thermodynamically and electromagnetically. In particular, the net radiative signal must vanish at global equilibrium, and the quantities used later as network weights must be derived from physical heat-flux sensitivities rather than introduced as unconstrained numerical coefficients.

For a harmonic mode of angular frequency $\omega$, the mean thermal energy of a bosonic oscillator is described by the Planck function

$$\Theta(\omega,T)=\frac{\hbar\omega}{\exp\left(\frac{\hbar\omega}{k_BT}\right)-1}. \tag{1}$$

Here $\hbar$ is the reduced Planck constant, $k_B$ is the Boltzmann constant, $\omega$ is angular frequency, and $T$ is absolute temperature. If the symmetrized oscillator energy is written as

$$E(\omega,T) = \frac{\hbar\omega}{2} + \Theta(\omega,T), \tag{2}$$

the zero-point contribution $\hbar\omega/2$ cancels from the difference between two bodies and therefore does not contribute to the net thermal heat flux.

For two bodies $i$ and $j$, define $\Theta_i(\omega) = \Theta(\omega,T_i)$ and $\Theta_j(\omega) = \Theta(\omega,T_j)$. The pairwise net radiative heat flux is written as [47,56]

$$Q_{i\to j} = \int_0^\infty \frac{d\omega}{2\pi}\left[\Theta_i(\omega) - \Theta_j(\omega)\right]\Phi_{ij}(\omega) \tag{3}$$

The sign convention used throughout this manuscript is $Q_{i\to j} > 0$ when the net radiative heat flux is directed from body $i$ toward body $j$. When $T_i = T_j$, the thermal occupation difference vanishes and $Q_{i\to j} = 0$, ensuring thermodynamic consistency.

For planar or locally planar structures, the spectral transmission function may be written as

$$\Phi_{ij}(\omega) = \sum_{p=s,p}\int \frac{d^2k_\parallel}{(2\pi)^2}\,\xi_{ij}^p(\omega,k_\parallel), \tag{4}$$

where $k_\parallel$ is the wavevector component parallel to the interfaces, $p$ denotes polarization, and $\xi_{ij}^p$ is the mode-resolved energy transmission coefficient. The vacuum wavenumber is $k_0 = \omega/c$. For a vacuum gap, the normal wavevector is $k_{z0} = \sqrt{k_0^2 - k_\parallel^2}$ for propagating modes. For evanescent modes with $k_\parallel > k_0$, $k_{z0} = i\kappa$ and $\kappa = \sqrt{k_\parallel^2 - k_0^2}$. The exponential factor $\exp(-2\kappa d_{ij})$ makes the evanescent contribution strongly dependent on the nanoscale gap $d_{ij}$.

For two planar half spaces, representative transmission factors can be written as

$$\xi_p^{\mathrm{prop}} = \frac{\left(1-\left|r_i^p\right|^2\right)\left(1-\left|r_j^p\right|^2\right)}{\left|1 - r_i^p r_j^p \exp\left(2ik_{z0}d_{ij}\right)\right|^2}, \tag{5}$$

$$\xi_p^{\mathrm{eva}} = \frac{4\,\mathrm{Im}\left(r_i^p\right)\mathrm{Im}\left(r_j^p\right)\exp\left(-2\kappa d_{ij}\right)}{\left|1 - r_i^p r_j^p \exp\left(-2\kappa d_{ij}\right)\right|^2}. \tag{6}$$

For a polar dielectric such as h-BN, the dielectric response can in general be tensorial. In the actual MATLAB implementation used for the present baseline simulations, however, the BN layer is represented by the same scalar Lorentz-oscillator dielectric function in the two polarization channels. The implemented model is $\varepsilon_{\mathrm{BN}}(E) = \varepsilon_0(E^2 - E_L^2 + iE\Gamma)/(E^2 - E_T^2 + iE\Gamma)$, with $\varepsilon_0 = 4.46$, $E_L = 0.1616$ eV, $E_T = 0.1309$ eV, and $\Gamma = 6.55\times10^{-4}$ eV. The more general tensorial treatment remains compatible with the theoretical framework but is not assumed in the numerical results reported here.

For a PCM-containing gate, the dielectric response depends on angular frequency, temperature, and material phase state:

$$\varepsilon_{\mathrm{PCM}} = \varepsilon_{\mathrm{PCM}}(\omega,T,f_{\mathrm{PCM}}), \tag{7}$$

where $f_{\mathrm{PCM}}$ is the dimensionless PCM phase fraction. The physical modulation chain is therefore

$$T \to f_{\mathrm{PCM}} \to \varepsilon_{\mathrm{PCM}} \to r^p \to \xi^p \to Q. \tag{8}$$

Patterned structures introduce additional geometrical dependence. In all simulations reported here, the one-dimensional PCM grating is treated with the second-order effective-medium approximation rather than RCWA. The filling ratio is defined as $f_g = w/\Lambda$, where $w$ is the PCM ridge width and $\Lambda$ is the grating period. The second-order

TE and TM effective permittivities follow the subwavelength-grating formulation of Chen, Zhang, and Timans [57], and the resulting effective dielectric response is then used in the multilayer reflection and transmission calculation.

For a multi-terminal node $i$, power balance is written as

$$C_i \frac{dT_i}{dt} = P_i^{\mathrm{ext}} + P_i^{\mathrm{rad}} + P_i^{\mathrm{cond}} - P_i^{\mathrm{loss}}, \tag{9}$$

where $C_i$ is thermal capacitance, $P_i^{\mathrm{ext}}$ is externally supplied power, $P_i^{\mathrm{rad}}$ is net radiative power received by the node, $P_i^{\mathrm{cond}}$ is conductive power when included, and $P_i^{\mathrm{loss}}$ represents other thermal losses. The radiative power is $P_i^{\mathrm{rad}} = \sum_{j \neq i} P_{j \to i}$. A heat-flux state is defined through the node reference area $A_i$ as $Q_i^{\mathrm{net}} = P_i^{\mathrm{rad}}/A_i$.

The normalized continuous thermal-information state is

$$h_i = \frac{Q_i^{\mathrm{net}}}{Q_{\mathrm{ref}}}, \tag{10}$$

where $Q_{\mathrm{ref}}$ is a chosen heat-flux normalization scale. The corresponding tri-state logic is

$$y_i = \mathcal{D}_Q\left(Q_i^{\mathrm{net}}; Q_{\mathrm{th}}^{\mathrm{L}}, Q_{\mathrm{th}}^{\mathrm{U}}\right) \tag{11}$$

Here $\mathcal{D}_Q$ is the physical tri-state decoder: $y_i = +1$ when $Q_i^{\mathrm{net}} > Q_{\mathrm{th}}^{\mathrm{U}}$, $y_i = 0$ when $Q_{\mathrm{th}}^{\mathrm{L}} \leq Q_i^{\mathrm{net}} \leq Q_{\mathrm{th}}^{\mathrm{U}}$, and $y_i = -1$ when $Q_i^{\mathrm{net}} < Q_{\mathrm{th}}^{\mathrm{L}}$. The finite interval between $Q_{\mathrm{th}}^{\mathrm{L}}$ and $Q_{\mathrm{th}}^{\mathrm{U}}$ gives the neutral state a physically meaningful tolerance rather than requiring the exact mathematical condition $Q = 0$.

To map radiative transport onto a reduced weighted operation, the local differential heat-flux conductance is defined as

$$G_{ij}^{Q} = \left.\frac{\partial Q_i^{\mathrm{net}}}{\partial T_j}\right|_{\mathrm{op}}. \tag{12}$$

Using $x_j = \delta T_j / T_{\mathrm{ref}}$, the corresponding dimensionless physical radiative weight is

$$W_{ij}^{\mathrm{rad}} = \frac{G_{ij}^{Q} T_{\mathrm{ref}}}{Q_{\mathrm{ref}}}. \tag{13}$$

Around an operating point, the preactivation becomes

$$u_i \approx \sum_j W_{ij}^{\mathrm{rad}} x_j + b_i^{\mathrm{th}}, \tag{14}$$

and the actual nonlinear device response produces the normalized heat-flux state $h_i = \mathcal{F}_i^{\mathrm{act}}(u_i)$. This establishes a consistent state hierarchy: $x$ is the normalized temperature input, $u$ is the normalized physical preactivation, $h$ is the physical normalized heat-flux output, $u^F$ is the fused preactivation used when multiple physical branches are combined, and $y$ is the decoded tri-state logic. Signed effective computational coefficients can be represented using two non-negative physical channels, $W_{ij}^{\mathrm{eff}} = W_{ij}^{\mathrm{pos}} - W_{ij}^{\mathrm{neg}}$, rather than interpreting a negative mathematical coefficient as a negative passive radiative conductance. This physically constrained representation is used later for signed convolution kernels.

## Near-Field Radiative Thermal Computing Primitives

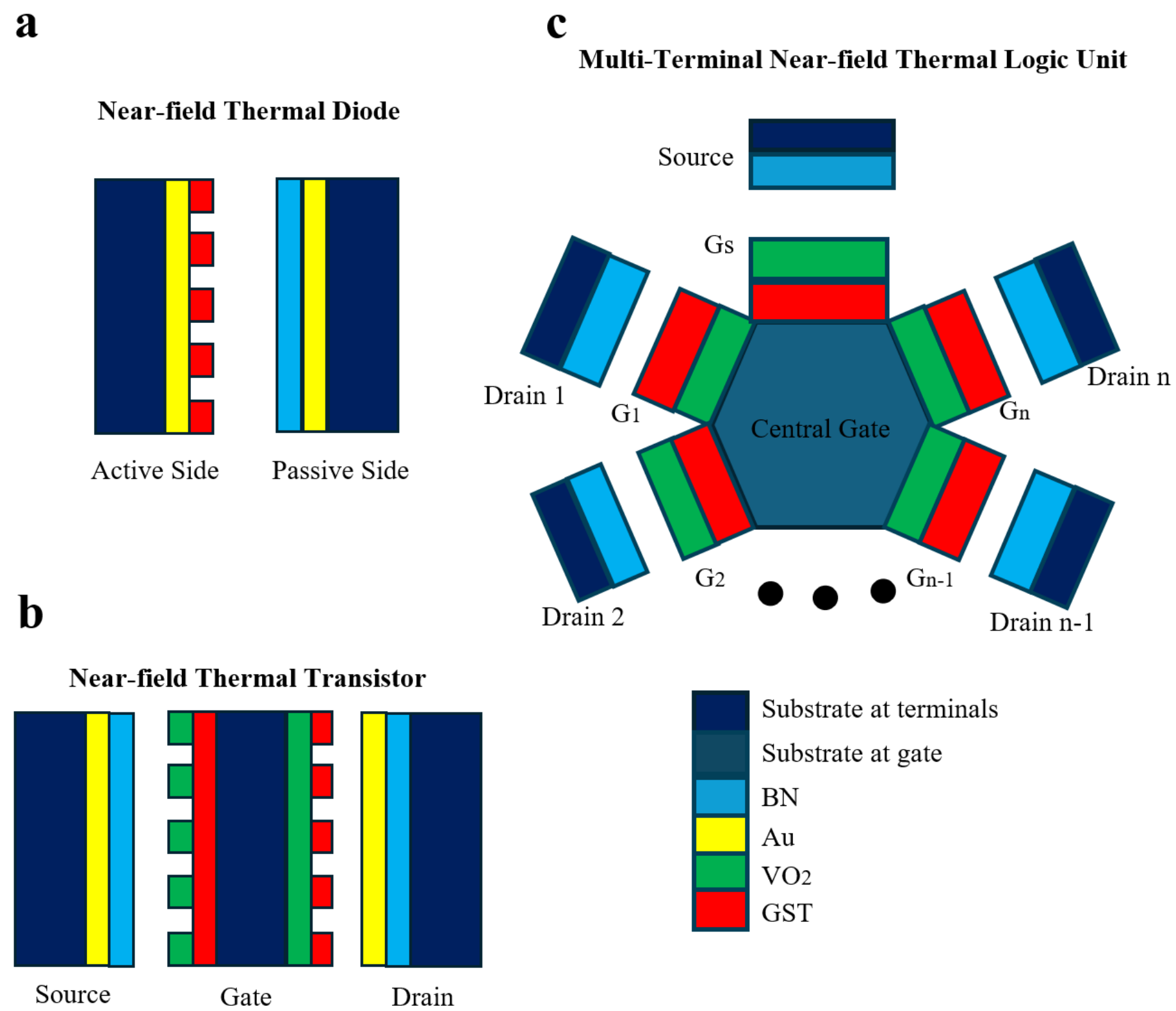

**Figure 1. Near-field radiative thermal computing primitives. (a) Near-field radiative thermal diode based on asymmetric material/structural response. (b) Source-gate-drain radiative thermal transistor/modulator containing phase-change materials. (c) Multi-terminal radiative thermal logic unit. The figure is a schematic derived from the device concepts used in Refs. 45 and 46.**

The near-field radiative thermal network is constructed from a small set of physical computing primitives rather than from abstract neurons introduced independently of the devices. The principal elements are a near-field radiative thermal diode, a gate-controlled radiative thermal transistor or modulator, and a multi-terminal thermal logic unit. Their respective roles are directional response, programmable nonlinear modulation, and fusion of several radiative inputs into a continuous or discrete output. [28,45,46] For a source-gate-drain unit, the physical temperatures are $T_S$, $T_G$, and $T_D$. The source supplies a thermal excitation, the drain provides the observable radiative response, and the gate modifies the electromagnetic transmission through its temperature- and phase-dependent optical state. In the simplest pairwise representation, the drain receives $P_D^{\mathrm{rad}} = P_{S\to D} + P_{G\to D}$, and the corresponding heat flux is $Q_D^{\mathrm{net}} = P_D^{\mathrm{rad}}/A_D$. A complete analysis remains subject to the steady or transient energy balance of the three-terminal structure.

The thermal diode is characterized by different heat-flux magnitudes under reversed thermal bias. If $Q_F = Q\left(T_{H,T_C}\right)$ is the forward heat flux and $Q_R = Q\left(T_{C,T_H}\right)$ is the reverse heat flux, one bounded rectification coefficient is

$$\mathcal{R}_D = \frac{|Q_F| - |Q_R|}{\max(|Q_F|, |Q_R|)} \tag{15}$$

The physical origin of rectification is the state dependence of the optical response: reversing the thermal bias can change the material state, reflection coefficients, electromagnetic transmission, and therefore the magnitude of radiative transfer. A reduced leaky directional model may be used only as an approximation to the calculated device response.

$$\mathcal{F}_D^{\text{act}}(u) = \begin{cases} u, & u \geq 0 \\ \eta_D u, & u < 0 \end{cases}, \qquad 0 \leq \eta_D < 1 \tag{16}$$

The parameter $\eta_D$ represents the reduced reverse response of the device model. If it is nonzero, the numerical derivative in the reverse branch is also nonzero. The gate-controlled thermal transistor is more carefully interpreted as a radiative thermal modulator unless genuine differential amplification is explicitly demonstrated. Its PCM gate obeys $\varepsilon_G = \varepsilon_G(\omega, T_{G,f_{\text{PCM}}})$, producing the physical sequence $T_G \to f_{\text{PCM}} \to \varepsilon_G \to r_G^p \to \xi \to Q_D^{\text{net}}$. A useful gate sensitivity is

$$\chi_G = \left.\frac{\partial Q_D^{\text{net}}}{\partial T_G}\right|_{\text{op}}. \tag{17}$$

A ratio such as $Q_{G\to D}/Q_{S\to D}$ is not by itself a transistor gain because an externally maintained gate can supply additional thermal power. If differential thermal amplification is evaluated, an appropriate quantity is $\alpha_T = |\partial Q_D^{\text{net}}/\,\partial Q_G^{\text{net}}|$, accompanied by a complete energy balance. The multi-terminal logic unit combines several thermal inputs through physically constrained radiative couplings. A normalized preactivation is $u_i = \sum_j W_{ij}^{\text{rad}}\, x_j + b_i^{\text{th}}$. Physical diode or transistor branches generate continuous outputs $h_i^{(r)}$. These can be fused as

$$u_i^F = \sum_r \gamma_r\, h_i^{(r)} + b_i^F, \tag{18}$$

where $u_i^F$ is the fused physical preactivation, $\gamma_r$ are fusion coefficients, and $b_i^F$ is a fusion offset. The physical tri-state output is still determined by the heat-flux thresholds, not by a SoftMax probability. When a differentiable numerical classifier is useful for optimization, logits $\ell_{i,c} = a_c\, u_i^F + b_c$ and SoftMax probabilities $\pi_{i,c} = \frac{\exp(\ell_{i,c})}{\sum_{c'} \exp(\ell_{i,c'})}$ can be introduced. These probabilities are numerical optimization variables rather than thermodynamic probabilities. The output can remain in the thermal domain through $Q_{i\to j}^{\text{out}}$, or it can be coupled to a detector plane for conventional readout. The device-level pathway is therefore temperature input → physical radiative coupling → directional or gate-controlled nonlinear response → continuous thermal fusion → heat-flux threshold decoding. This sequence establishes the building blocks used in the spatial and temporal network architectures.

**Table 1. Physical computing primitives and their reduced network roles.**

| Primitive | Primary physical mechanism | Measured/controlled quantity | Reduced network role |
|---|---|---|---|
| Thermal source/reservoir | Sets thermal occupation and radiative driving force | $T_S$, $P_{\text{ext}}$ | Input/bias state |
| Near-field thermal diode | Bias-dependent asymmetric radiative transmission | $Q_F$, $Q_R$ | Directional nonlinearity/rectification |
| Gate-controlled radiative transistor/modulator | PCM- and temperature-dependent modulation of transmission | $T_G$, $f_{\text{PCM}}$, $Q_D$ | Programmable nonlinear coupling |
| Multi-terminal logic unit | Combination of several radiative inputs | $Q_i^{\text{net}}$, $u_i^F$ | Fusion and tri-state decoding |
| Thermal storage node | Finite heat capacity and thermal relaxation | $s_i$, $\tau_i^{\text{th}}$ | Volatile recurrent memory |
| PCM memory state | Hysteresis or persistent phase state | $f_{\text{PCM}}$, $\mathcal{H}_i$ | History-dependent/potential nonvolatile memory |
| Detector/screen | Thermal-to-observable transduction | $Q_{\text{out}}$, $T_{\text{det}}$, detector signal | Output/readout |

## Physical Mapping from Radiative Transport to Neural Operations

The device primitives can be reduced to operations resembling weighted summation, biasing, nonlinear transformation, and fusion, but the reduced variables must remain connected to the underlying transport physics. The net heat flux of node $i$ is a nonlinear function of temperatures, optical properties, PCM states, gaps, and geometry: $Q_i^{\text{net}} = Q_i(T, \varepsilon, f_{\text{PCM},d}, \theta_{geo})$. Around a selected operating state, the response can be linearized in the physical variables. The physical programming/design parameter vector is denoted consistently by $\theta$ and may contain gate temperature, PCM phase fraction, gap distance, grating filling ratio, PCM thickness, and other controllable parameters. The local temperature contribution is governed by $G_{ij}^{Q} = (\partial Q_i^{\text{net}}/\partial T_j)|_{op}$. With $x_j = \delta T_j/T_{\text{ref}}$ and $h_i = \delta Q_i/Q_{\text{ref}}$, the first-order weighted relation is

$$u_i = \sum_j W_{ij}^{\text{rad}}\, x_j + b_i^{\text{th}} \tag{19}$$

The weight $W_{ij}^{\text{rad}}$ is physically realizable only through an admissible parameter set $\boldsymbol{\theta} \in \Omega_{\text{phys}}$. A target numerical weight $W_{ij}^{\text{tar}}$ can be implemented exactly only if it lies inside the physically accessible set. Otherwise, the programming problem is

$$\theta_{ij}^{\text{opt}} = \text{argmin}_{\theta_{ij}\in\Omega_{\text{phys}}}\mathcal{E}_{ij}^{\text{map}} \tag{20}$$

where $\mathcal{E}_{ij}^{\text{map}} = \left|W_{ij}^{\text{rad}}(\theta_{ij}) - W_{ij}^{\text{tar}}\right|^2$. This gives the sequence $\mathbf{W}^{\text{tar}} \to \boldsymbol{\theta}^{\text{opt}} \to \mathbf{W}^{\text{rad}}$ and distinguishes the numerical design objective from the actual radiative coupling produced by the structure. The physically accessible weight range, resolution, monotonicity, uncertainty, and cross-dependence are important characteristics. Sensitivities such as $\partial W^{\text{rad}}/\partial T_G$, $\partial W^{\text{rad}}/\partial d$, and $\partial W^{\text{rad}}/\partial f_{\text{PCM}}$ quantify programmability and robustness. PCM hysteresis can make $W^{\text{rad}} = W^{\text{rad}}(T_{G,\mathcal{H}})$, so the same gate temperature can correspond to different couplings depending on thermal history. Signed mathematical weights are represented differentially as $W_{ij}^{\text{eff}} = W_{ij}^{\text{pos}} - W_{ij}^{\text{neg}}$ using two non-negative physical branches. The nonlinear response is $h_i = \mathcal{F}_i^{\text{act}}(u_i, s_i)$, where the physical activation is obtained from the actual device response. ReLU-like, sigmoid-like, or leaky analytical forms are reduced approximations unless they are explicitly fitted to the calculated thermal response.

For several branches, the fused physical preactivation is $u_i^F = \sum_r \gamma_r\, h_i^{(r)} + b_i^F$. The decoded physical logic $y_i$ is obtained from heat-flux thresholds. A separate numerical SoftMax head may be used during optimization but is not the physical logic mechanism. At larger perturbations, higher-order terms become important: $\delta Q_i = \sum_j G_{ij}^{Q}\,\delta T_j + (1/2)\sum_{j,k} H_{ijk}^{Q}\,\delta T_j \delta T_k + \cdots$. Consequently, the physical weights are generally state dependent over a sufficiently large temperature range. A selected operating interval is therefore required if the linearized weight interpretation is to remain accurate. Numerical learning is interpreted as physical-parameter optimization. The chain rule $\partial L/\partial\theta_r = \sum_{ij}(\partial L/\partial W_{ij}^{\text{rad}})(\partial W_{ij}^{\text{rad}}/\partial\theta_r)$ translates a computational objective into a desired physical parameter change. A projected update $\theta^{n+1} = \Pi_{\Omega_{\text{phys}}}[\theta^n - \alpha_{\text{opt}}\nabla_\theta L]$ keeps the programmed state physically admissible. Backpropagated gradients are numerical sensitivities and not reverse physical heat currents.

## Physical Radiative Thermal Convolutional Network

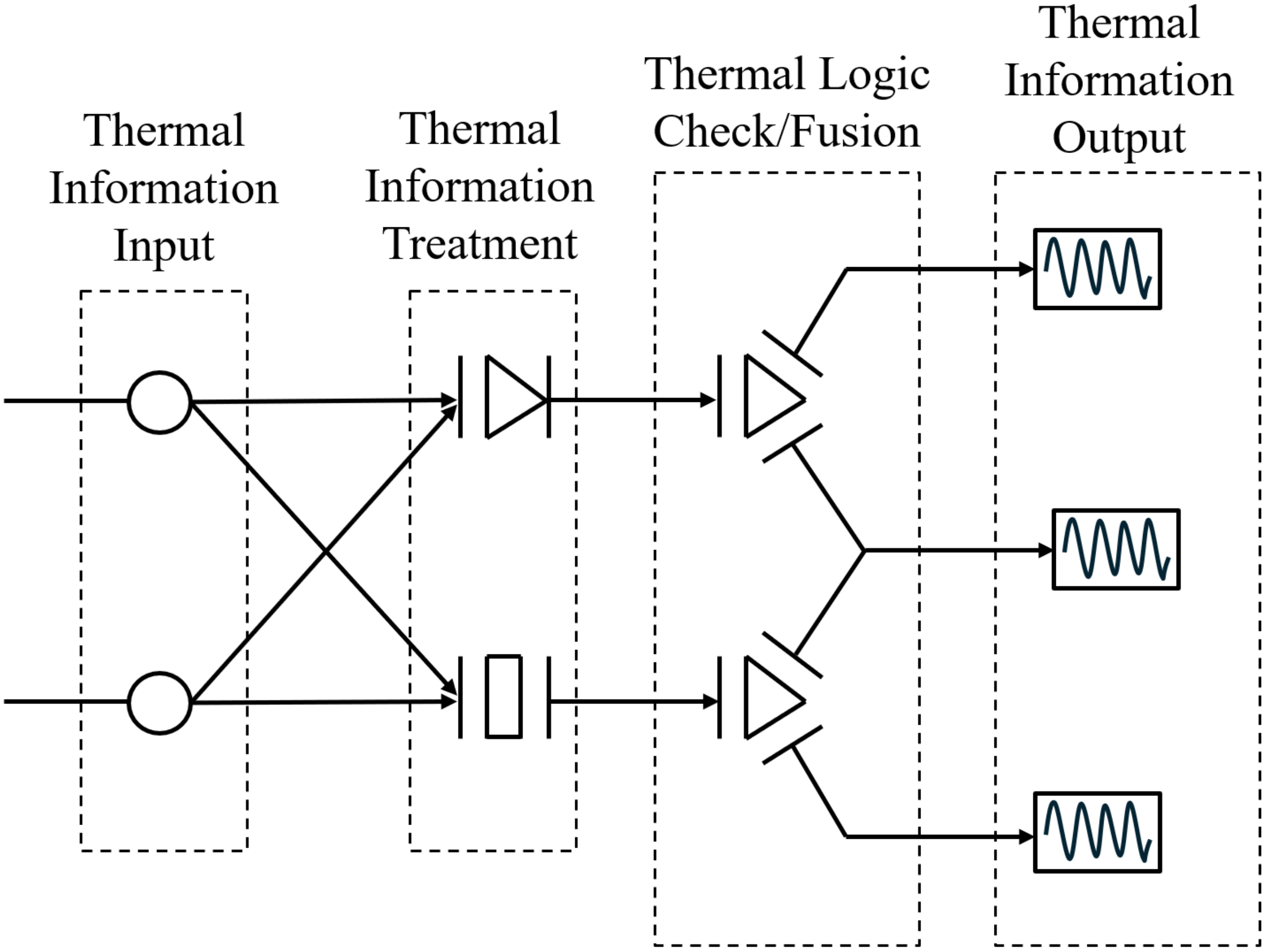

**Figure 2. Functional architecture of the physical radiative T-CNN. The four functional stages are thermal-information input, local radiative treatment, thermal logic check/fusion, and thermal-information output. The convolutional interpretation applies specifically where a finite local radiative kernel is translated across a spatial input; the schematic is conceptual rather than a fabricated multilayer chip.**

A physical convolution requires more than weighted summation. It requires a finite local receptive field, a spatially defined kernel, translation of that kernel across an input field, approximate sharing of the same coupling pattern at different positions, defined stride and boundary conditions, and formation of a spatial output feature map. The T-CNN is therefore defined as a spatial radiative thermal network in which a local temperature field is mapped through a physically realizable radiative kernel into a heat-flux feature field.

$$x_{m,n}^{(0)} = \frac{T_{m,n}^{(0)} - T_0^{(0)}}{T_{\mathrm{ref}}}. \tag{21}$$

For an output node (p,q), a finite receptive field contains the positions $(pS_x + u, qS_y + v)$, where $S_x$ and $S_y$ are spatial strides and u,v are local kernel offsets. Around the operating point, the local heat-flux response is

$$\delta Q_{p,q}^{(l)} \approx \sum_{(u,v)\in\mathcal{N}_{p,q}} G_{p,q;u,v}^{Q,(l)} \; \delta T_{pS_x+u,\, qS_y+v}^{(l-1)} \tag{22}$$

The dimensionless physical kernel coefficient is

$$K_{p,q;u,v}^{(l),\mathrm{rad}} = \frac{G_{p,q;u,v}^{Q,(l)} T_{\mathrm{ref}}}{Q_{\mathrm{ref}}} \tag{23}$$

For an ideal translationally invariant layer, $K^{\mathrm{rad}}_{p,q;u,v} \approx K^{\mathrm{rad}}_{u,v}$. The convolutional preactivation field is

$$u^{(l)}_{p,q} = \sum_u \sum_v K^{(l),\mathrm{rad}}_{u,v}\, x^{(l-1)}_{pS_x+u,\, qS_y+v} + b^{(l),\mathrm{th}}_{p,q} \tag{24}$$

followed by the physical nonlinear response $h^{(l)}_{p,q} = \mathcal{F}^{(l),\mathrm{act}}_{p,q}\left(u^{(l)}_{p,q}\right)$. The output feature map is $H^{((l))} = \left[h^{((l))}_{p,q}\right]$.

Boundary conditions must be defined explicitly. A valid convolution evaluates only positions whose receptive fields lie fully inside the physical array. A zero-padding approximation corresponds to zero normalized perturbation, not zero absolute temperature. A more physical edge model introduces a boundary reservoir temperature $T_B$ and corresponding $x_B = (T_B - T_0)/T_{\mathrm{ref}}$. Physical weight sharing is approximate rather than exact. A local kernel can be written $K^{\mathrm{rad}}_{p,q;u,v} = K^{\mathrm{rad}}_{u,v} + \delta K_{p,q;u,v}$, and a sharing error can quantify deviations across the array. Likewise, radiative cross-talk outside the intended receptive field must remain small compared with the desired local coupling if the convolutional interpretation is to hold. The target mathematical kernel $\mathbf{K}^{\mathrm{tar}}$ and physical kernel $\mathbf{K}^{\mathrm{rad}}$ are compared through $\mathcal{E}_K = \| \mathbf{K}^{\mathrm{rad}} - \mathbf{K}^{\mathrm{tar}} \|_F / (\| \mathbf{K}^{\mathrm{tar}} \|_F + \delta_{\mathrm{num}})$. The target and physical feature maps are similarly compared using $\mathcal{E}_{\mathrm{map}} = \| \mathbf{H}^{\mathrm{rad}} - \mathbf{H}^{\mathrm{tar}} \|_F / (\| \mathbf{H}^{\mathrm{tar}} \|_F + \delta_{\mathrm{num}})$. These two metrics provide direct tests of physical kernel fidelity and computational output fidelity.

A signed edge kernel such as

$$\mathbf{K}^{\mathrm{tar}}_x = \begin{bmatrix} -1 & 0 & +1 \\ -1 & 0 & +1 \\ -1 & 0 & +1 \end{bmatrix} \tag{25}$$

is implemented differentially by decomposing it into positive and negative physical branches $\mathbf{K}^{\mathrm{tar}}_x = \mathbf{K}^{\mathrm{pos}}_x - \mathbf{K}^{\mathrm{neg}}_x$. The subtraction must be implemented by a physical differential stage or declared explicitly as numerical post-processing. A key dimensional issue arises when multiple physical layers are cascaded. The output $h$ of one layer is a normalized heat flux, whereas the input $x$ of the next layer is a normalized temperature. Under a local steady-state thermal-resistance approximation, $\delta T_i \approx R_{\mathrm{th},i}\, A_i\, \delta Q_i$. Therefore

$$x^{(l+1)}_i = \beta^{(l)}_i h^{(l)}_i, \qquad \beta^{(l)}_i = \frac{R^{(l)}_{\mathrm{th},i} A^{(l)}_i Q_{\mathrm{ref}}}{T_{\mathrm{ref}}}. \tag{26}$$

This explicit heat-flux-to-temperature transduction prevents the output of one layer from being silently reinterpreted as the next layer's temperature input. In a dynamic network, the steady relation is replaced by the node energy balance. The original four-stage schematic is most safely interpreted as thermal input → local radiative treatment → logic fusion → thermal output. The convolution occurs only in the stage where a finite shared radiative kernel is translated across the spatial field. An explicit target-versus-radiative kernel demonstration is therefore necessary if the term T-CNN is retained.

## Learning and Optimization of the Physical T-CNN

The physical T-CNN performs forward spatial processing through radiative kernels and nonlinear thermal-device responses. The optimization procedure determines which physically admissible parameters are programmed so that the network approximates a desired operation. Numerical optimization is therefore distinguished from intrinsic physical learning.

$$\boldsymbol{\theta}^{\mathrm{opt}} = \arg\min_{\boldsymbol{\theta} \in \Omega_{\mathrm{phys}}} \mathcal{L}_{\text{T-CNN}}(\boldsymbol{\theta}) \tag{27}$$

For tri-state classification, the numerical target at node $i$ is a one-hot vector $t_i$ over $c \in (-1, 0, +1)$. A class-balanced cross-entropy is

$$\mathcal{L}_{\mathrm{CE}} = -\frac{1}{N_s} \sum_{i,c} w_c\ t_{i,c} \ln\left(\pi_{i,c} + \delta_{\mathrm{num}}\right) \tag{28}$$

For a one-hot target with correct class $c^{\mathrm{true}}_i$, the SoftMax-cross-entropy derivative is

$$\frac{\partial \mathcal{L}_{\text{CE}}}{\partial \ell_{i,k}} = \frac{w_{c_i^{\text{true}}}}{N_s}\left(\pi_{i,k} - t_{i,k}\right) \tag{29}$$

The classification term does not by itself guarantee physical robustness. A thermal margin penalty can instead use the heat-flux distance from the required threshold. For positive, negative, and neutral targets, the margin is defined relative to $Q_{\text{th}}^{\text{U}}$, $Q_{\text{th}}^{\text{L}}$, or the nearest neutral boundary, respectively. A desired minimum margin $M_Q^{req}$ can be enforced with a squared hinge penalty.

A spatial smoothness loss may penalize unsupported node-to-node oscillation:

$$\mathcal{L}_{\text{smooth}} = \frac{1}{N_{\text{edge}}}\sum_{m,n}\left[\left(h_{m+1,n} - h_{m,n}\right)^2 + \left(h_{m,n+1} - h_{m,n}\right)^2\right] \tag{30}$$

Other physically motivated terms may include radiative-weight regularization, target-kernel mismatch, feature-map mismatch, physical-bound penalties, and cross-talk penalties. To avoid conflict with the retention factor $\lambda_{\text{ret}}$ used in the T-RNN, numerical loss weights are denoted here by $\rho$ rather than plain $\lambda$.

$$\begin{aligned}\mathcal{L}_{\text{T-CNN}} = \quad & \rho_{\text{CE}}\mathcal{L}_{\text{CE}} + \rho_{\text{margin}}\mathcal{L}_{\text{margin}} + \rho_{\text{smooth}}\mathcal{L}_{\text{smooth}} \\ & +\rho_W\mathcal{L}_W + \rho_K\mathcal{L}_K + \rho_H\mathcal{L}_H\end{aligned} \tag{31}$$

The numerical gradient propagates through the fused state, physical response model, and radiative kernel. For a kernel coefficient, $\partial u_{p,q} / \partial K_{u,v}^{\text{rad}} = x_{pS_x+u,qS_y+v}$. The chain rule then maps the kernel gradient to physical parameters through $\partial K^{\text{rad}} / \partial \theta_r$.

The physically meaningful update is a projected update within $\Omega_{\text{phys}}$:

$$\boldsymbol{\theta}^{(n+1)} = \Pi_{\Omega_{\text{phys}}}\left[\boldsymbol{\theta}^{(n)} - \alpha_{\text{opt}}\nabla_{\boldsymbol{\theta}}\mathcal{L}\right]. \tag{32}$$

PCM hysteresis makes the programming trajectory important because a desired phase fraction cannot always be assigned independently of thermal history. The optimizer therefore determines a desired physical state, while the actual device must reach that state through an admissible heating or cooling path. The most appropriate learning architecture for the present framework is therefore offline numerical optimization followed by programmed physical inference. Optimization convergence and thermal equilibrium remain separate concepts: the former refers to stabilization of a numerical objective, whereas the latter requires the physical energy-balance residual to approach zero.

**Physical Radiative Thermal Recurrent Network**

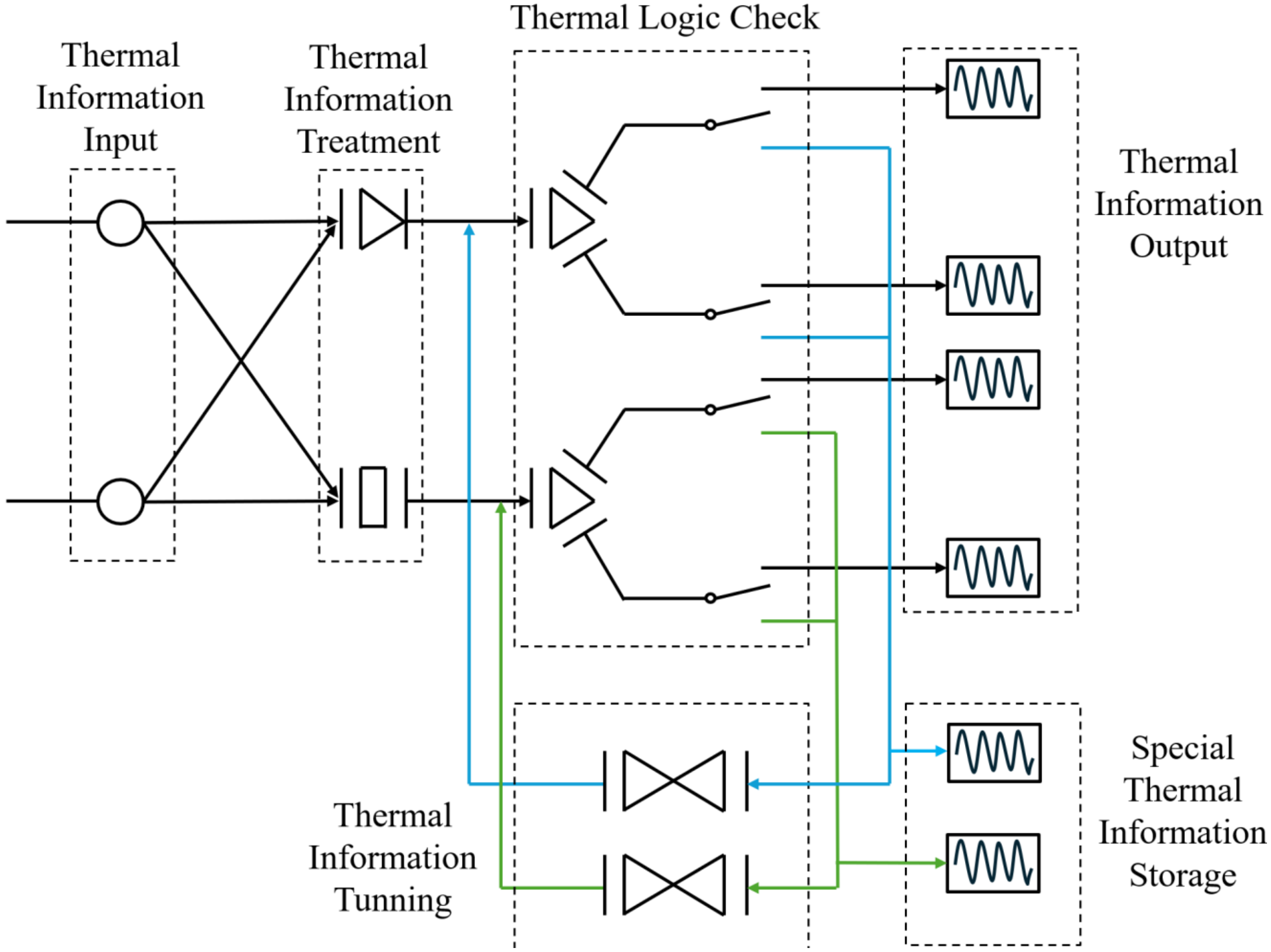

**Figure 3. Conceptual physical radiative T-RNN. Recurrent pathways reintroduce previous thermal outputs through radiative/thermal feedback while PCM state evolution and finite thermal relaxation provide history dependence. The recurrent memory is interpreted through the coupled energy-balance and phase-state equations rather than as a direct physical implementation of software LSTM/GRU gates.**

A physical T-RNN must contain a state that persists and influences later thermal responses. Its recurrence is derived from finite thermal capacitance, radiative feedback, and PCM state evolution rather than from the algebraic structure of a software RNN.

$$C_i\dot{T}_i = P_i^{\mathrm{ext}} + P_i^{\mathrm{rad}} + P_i^{\mathrm{cond}} - P_i^{\mathrm{loss}} \tag{33}$$

The present temperature contains the integral of the previous net thermal power, so finite thermal capacitance provides the first physical source of recurrence. The PCM adds a second state variable governed generally by

$$\frac{df_{\mathrm{PCM},i}}{dt} = \mathcal{K}_i\left[T_i, f_{\mathrm{PCM},i}, \frac{dT_i}{dt}, \mathcal{H}_i\right]. \tag{34}$$

Because $\varepsilon_i = \varepsilon_i\left(\omega, T_{i,f_{\mathrm{PCM},i}}\right)$, the material state modifies reflection, transmission, heat flux, and therefore the next temperature. The physical recurrent loop is

$$T(t) \to f_{\mathrm{PCM}}(t) \to \varepsilon(t) \to r^p(t) \to \xi^p(t) \to Q(t) \to T(t+\Delta t) \tag{35}$$

A reduced volatile thermal-storage state can be derived from the linearized energy balance. If $\delta P_i^{\mathrm{in}} = A_i\, Q_{\mathrm{ref}}\, h_i$ and $G_i^{\mathrm{th}}$ is the effective relaxation conductance,

$$\tau_i^{\mathrm{th}}\frac{ds_i}{dt} = -s_i + \eta_{\mathrm{store},i} h_i, \tag{36}$$

$$\tau_i^{\mathrm{th}} = \frac{C_i}{G_i^{\mathrm{th}}}, \qquad \eta_{\mathrm{store},i} = \frac{A_i Q_{\mathrm{ref}}}{G_i^{\mathrm{th}} T_{\mathrm{ref}}}. \tag{37}$$

For a constant input over one time interval $\Delta t$,

$$\begin{aligned} s_{i,k+1} = \quad & \lambda_{\mathrm{ret},i} s_{i,k} \\ & + (1 - \lambda_{\mathrm{ret},i}) \eta_{\mathrm{store},i} h_{i,k} \end{aligned} \tag{38}$$

$$\lambda_{\mathrm{ret},i} = \exp\left(-\frac{\Delta t}{\tau_i^{\mathrm{th}}}\right). \tag{39}$$

Thus $\lambda_{\mathrm{ret}}$ is the discrete physical retention factor, while $\tau_{\mathrm{th}}$ is the underlying physical relaxation time. A value $\lambda_{\mathrm{ret}}$ near one corresponds to strong retention from one sequence step to the next; a value near zero corresponds to rapid forgetting. The PCM introduces a separate material timescale $\tau_{\mathrm{PCM}}$. A more complete phase-changing node may also include latent heat, $C_i^{\mathrm{sens}} dT_i/dt + M_{\mathrm{PCM},i}\, L_i^{\mathrm{lat}}\, df_{\mathrm{PCM},i}/dt = P_i^{\mathrm{ext}} + P_i^{\mathrm{rad}} + P_i^{\mathrm{cond}} - P_i^{\mathrm{loss}}$. At discrete time $k$, define the feedforward contribution $u_{i,k}^{\mathrm{ff}} = \sum_j W_{ij}^{\mathrm{ff,rad}} x_{j,k}$, the recurrent contribution $u_{i,k}^{\mathrm{rec}} = \sum_r W_{ir}^{\mathrm{rec,rad}} \beta_r^{\mathrm{fb}} h_{r,k-1}$, and the retained-state contribution $u_{i,k}^{\mathrm{store}} = \sum_r W_{ir}^{\mathrm{store,rad}} s_{r,k}$. The total physical preactivation is

$$u_{i,k} = u_{i,k}^{\mathrm{ff}} + u_{i,k}^{\mathrm{rec}} + u_{i,k}^{\mathrm{store}} + b_i^{\mathrm{th}} \tag{40}$$

The physical nonlinear output is $h_{i,k} = \mathcal{F}_i^{\mathrm{act}}(u_{i,k}, T_{G,i,k}, f_{\mathrm{PCM},i,k})$, followed by branch fusion $u_{i,k}^F$ and heat-flux threshold decoding $y_{i,k}$. The first-order continuous memory kernel derived from the thermal balance is $M_i^{\mathrm{th}}(\Delta t) = (1/\tau_i^{\mathrm{th}})\exp(-\Delta t/\tau_i^{\mathrm{th}})$ for $\Delta t \geq 0$. More complex devices may require a sum of relaxation modes. $VO_2$ and GST must be assigned different memory semantics. $VO_2$ can provide history-dependent or hysteretic memory because the same present temperature may correspond to different phase fractions on heating and cooling branches. GST can potentially provide nonvolatile phase-state storage, but programming energy, switching kinetics, drift, and endurance must then be included. Thermal inertia, $VO_2$ hysteresis, and GST persistent phase storage are therefore distinct physical mechanisms. [50-52] The physical T-RNN is also distinct from the software LSTM/GRU used later. The physical recurrent state is described by temperatures, heat fluxes, thermal storage, and PCM phase, whereas the software model uses numerical hidden states and trainable matrices.

## Temporal Learning and Thermal Memory Characteristics

The recurrent network can be interpreted through a physical memory cycle: write → retain → read → update/reset. The effective memory state may depend on the retained thermal state $s_i$, PCM phase fraction $f_{\mathrm{PCM},i}$, and thermal history $\mathcal{H}_i$. This memory state is not a software hidden state. For a volatile thermal write pulse of amplitude $h_{i,w}$ and duration $t_w$, the first-order storage model gives

$$\begin{aligned} s_i(t_w) = \quad & s_i(0) e^{-t_w/\tau_i^{\mathrm{th}}} \\ & + \eta_{\mathrm{store},i} h_{i,w} \left(1 - e^{-t_w/\tau_i^{\mathrm{th}}}\right) \end{aligned} \tag{41}$$

The write energy is $E_i^{\mathrm{write}} = \int P_i^{\mathrm{write}}(t)dt$, and a reset operation similarly requires $E_i^{reset} = \int P_i^{reset}(t)dt$. Nonvolatile storage therefore does not imply zero programming energy. After the write signal is removed, the normalized volatile retention is

$$\mathcal{M}_i^{\mathrm{th}}(\Delta t) = \frac{s_i(t_w + \Delta t)}{s_i(t_w)} = \exp\left(-\frac{\Delta t}{\tau_i^{\mathrm{th}}}\right) \tag{42}$$

If $\zeta_{\min}$ is the minimum accepted retention fraction, the corresponding retention time is $t_{\mathrm{ret},i} = -\tau_i^{\mathrm{th}} \ln \zeta_{\min}$. In discrete time, $\mathcal{M}_i^{\mathrm{th}}(k\Delta t) = \lambda_{\mathrm{ret},i}^k$, so physical retention time and sequence memory depth can be distinguished. Readout occurs when the retained state modifies a physically coupled thermal node. For two memory states a and b, a normalized readout contrast is $C_{\mathrm{read}} = |Q_{\mathrm{read}}^a - Q_{\mathrm{read}}^b| / Q_{\mathrm{ref}}$. A non-destructive memory additionally requires small read disturbance relative to the stored state.

For $VO_2$, the strongest demonstration of hysteretic memory is an equal-temperature history test. Two thermal histories A and B satisfy $T_A(t_r) = T_B(t_r) = T_r$ but $f_{VO2}^A(t_r) \neq f_{VO2}^B(t_r)$, producing $Q_D^A(t_r) \neq Q_D^B(t_r)$. The normalized history separation is $M_H = |Q_D^A - Q_D^B|/Q_{\text{ref}}$. For GST, the same generic phase variable $f_{\text{PCM}}$ is retained to avoid introducing a conflicting $x$ symbol. A persistent programmed state requires $|df_{\text{PCM}}/dt| \approx 0$ over the specified retention interval. Any claim of nonvolatile operation is accompanied by write/reset conditions, drift, endurance, and retention criteria. The useful memory timescale is matched to the task timescale rather than maximized indefinitely. A previous state contributes meaningfully to the next input only if its retention over $\Delta t_{\text{in}}$ remains sufficiently large. Excessively short memory is ineffective; excessively long memory can cause unwanted carryover of obsolete states. Temporal optimization remains numerical. A sequence objective may combine sequence cross-entropy, physical margin, temporal smoothness, storage-dynamics consistency, and radiative-weight regularization:

$$\begin{aligned}\mathcal{L}_{\text{T-RNN}} = \quad & \rho_{\text{seq}}\mathcal{L}_{\text{seq}} + \rho_{\text{margin}}\mathcal{L}_{\text{margin}} + \rho_{\text{temp}}\mathcal{L}_{\text{temp}} \\ & + \rho_{\text{store}}\mathcal{L}_{\text{store}} + \rho_W\mathcal{L}_W\end{aligned} \tag{43}$$

The derivative $\partial s_{k+n}/\partial s_k = \lambda_{\text{ret}}^n$ provides a direct connection between physical retention and the numerical propagation of sensitivity through the reduced recurrent model. However, backpropagation through time remains a numerical algorithm and not a physical thermal signal.

## Proof-of-Principle Numerical Demonstrations of Radiative Thermal Computation

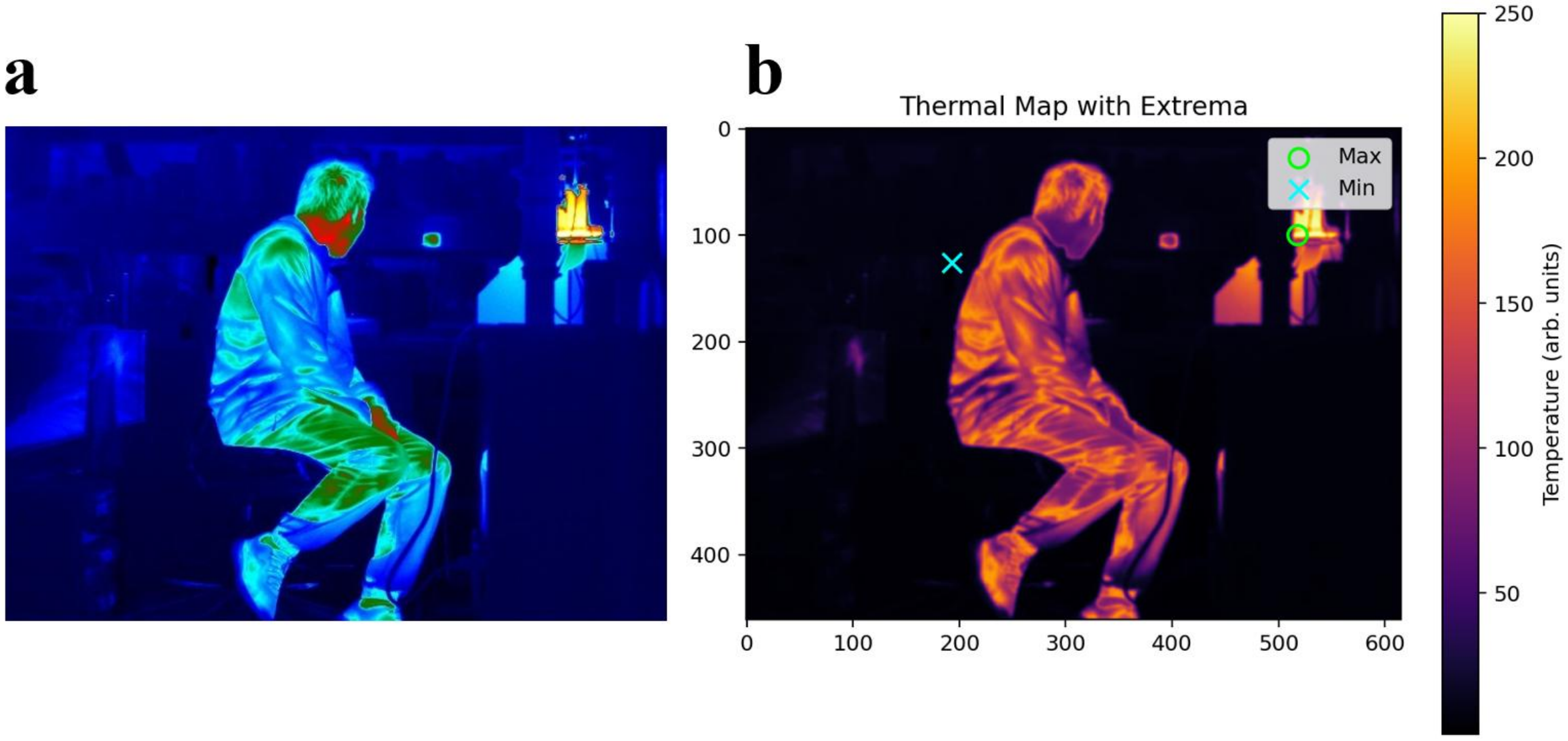

**Figure 4. Illustrative thermal-image extrema post-processing. Panel (a) is the input image supplied to the Python demonstration. Panel (b) is the post-processed output generated by the thermal-extrema routine. A Gaussian smoothing operation is applied before the global maximum and minimum are identified and overlaid.**

The framework is supported by numerical tests that demonstrate the operations attributed to the physical network. Three tests are especially important: conversion of a thermal field into tri-state radiative logic, an explicit target-versus-physical convolution, and a recurrent history-dependence test at equal present temperature. For a spatial thermal input $T_{\text{in}} = [T_{m,n}^{\text{in}}]$, the normalized field is $x_{m,n}^{\text{in}} = (T_{m,n}^{\text{in}} - T_0)/T_{\text{ref}}$. The physical calculation produces $Q_{m,n}^{\text{net}}$, from which $h_{m,n} = Q_{m,n}^{\text{net}}/Q_{\text{ref}}$ and the tri-state field $y_{m,n}$ are obtained. The basic information chain is

$$\mathbf{T}_{\text{in}} \rightarrow \mathbf{X}_{\text{in}} \rightarrow \mathbf{Q}_{\text{net}} \rightarrow \mathbf{H} \rightarrow \mathbf{Y}. \tag{44}$$

The robustness of the classified field can be quantified by the heat-flux margin $M_{Q,m,n}$ to the nearest required threshold, with $M_Q^{min} = min_{m,n} M_{Q,m,n}$. If an uncertainty $\sigma_Q$ is later available, $M_Q^{min}/\sigma_Q$ provides a direct logic robustness ratio. The essential T-CNN test is an explicit convolution. A target kernel $\mathbf{K}^{\text{tar}}$ is mapped onto a physically attainable $\mathbf{K}^{\text{rad}}$, followed by comparison of the target and physical feature maps. A representative target is the $3 \times 3$ horizontal-gradient kernel $\mathbf{K}_x^{\text{tar}}$. The physical signed operation is realized by positive and negative radiative branches rather than negative passive conductances.

$$\mathcal{E}_K = \frac{\| \mathbf{K}^{\text{rad}} - \mathbf{K}^{\text{tar}} \|_F}{\| \mathbf{K}^{\text{tar}} \|_F + \delta_{\text{num}}} \tag{45}$$

$$\mathcal{E}_{\text{map}} = \frac{\| \mathbf{H}^{\text{rad}} - \mathbf{H}^{\text{tar}} \|_F}{\| \mathbf{H}^{\text{tar}} \|_F + \delta_{\text{num}}} \tag{46}$$

The essential T-RNN test prepares two different thermal histories that arrive at the same read temperature:

$$\begin{aligned} T_G^A(t_r) &= T_G^B(t_r) = T_r, \\ f_{\text{PCM}}^A(t_r) &\neq f_{\text{PCM}}^B(t_r), \qquad Q_D^A(t_r) \neq Q_D^B(t_r) \end{aligned} \tag{47}$$

The phase-state separation $M_f = \left| f_{\text{PCM}}^A - f_{\text{PCM}}^B \right|$ and radiative history separation $M_H = |Q_D^A - Q_D^B| / Q_{\text{ref}}$ quantify the physical memory effect. A delayed write-read test can further define a radiative retention ratio relative to equilibrium. Each numerical result satisfy thermodynamic consistency: pairwise net heat flux vanishes at equal temperatures, and the implemented multi-terminal power balance is evaluated consistently with the adopted sign convention.

## Software-Based Recurrent Inverse Identification of Radiative Structural Parameters

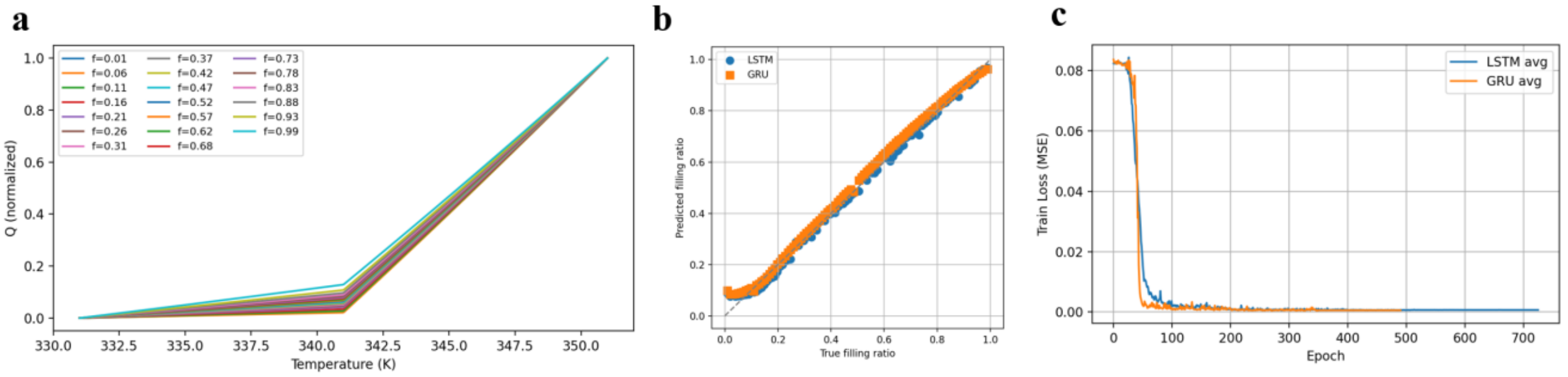

**Figure 5. Software-based inverse identification of the grating filling ratio from (a) simulated near-field radiative Q-T curves using (b) bidirectional LSTM and GRU models. And (c) shows the training loss during each epoch. These software results demonstrate structural information encoded in the simulated thermal response and are not a physical implementation of the radiative T-RNN.**

The inverse-identification study is conceptually separate from the physical T-RNN. It uses conventional software LSTM and GRU architectures to analyze precomputed near-field heat-flux–temperature curves and infer a structural parameter. Successful inverse prediction therefore demonstrates information content in the simulated radiative response, not physical implementation of an LSTM or GRU by the thermal device.

$$f_g \overset{\text{NFRHT}}{\rightarrow} Q(T; f_g) \overset{\text{LSTM/GRU}}{\rightarrow} f_g^{\text{pred}} \tag{48}$$

The final software dataset used by the saved PyTorch training run contains 99 independently labelled filling-ratio curves spanning approximately 0.01 to 0.99. Each curve contains 500 uniformly ordered temperature samples between 331 K and 351 K, giving 49,500 temperature-heat-flux samples in total. The statistically independent structural sample count is therefore 99 complete curves rather than the number of individual temperature points. Each curve is converted to three software input channels: min-max-normalized heat flux, the first numerical gradient of that normalized sequence, and the numerical gradient of the first-gradient sequence. The gradients are calculated with NumPy gradient along the ordered, uniformly sampled temperature sequence and are used as shape descriptors rather than as independently measured physical observables.

$$\mathbf{v}_k = \left[ \tilde{Q}_k \quad \tilde{Q}'_{T,k} \quad \tilde{Q}''_{T,k} \right]^{\mathsf{T}}. \tag{49}$$

The normalization is performed independently for each heat-flux curve using its own minimum and maximum values, with a numerical denominator offset of 1e-9. The derivative endpoint treatment follows the default NumPy gradient implementation. The inverse benchmark retains the idealized sharp VO2 state switch at 341 K used by the MATLAB dataset generator; this software-dataset assumption is kept separate from the finite-width and hysteretic phase-transition descriptions used in the physical recurrent-network discussion.

The software predictors use a single bidirectional LSTM or GRU layer with 32 hidden units per direction. Because the recurrent module contains one layer, PyTorch applies zero internal recurrent-layer dropout; the specified dropout probability of 0.15 is instead applied after temporal mean pooling and within the regression head. The pooled representation is passed through a linear layer, GELU activation, dropout, and a final linear layer, followed by a sigmoid output constraint. Training uses AdamW with an initial learning rate of 0.001, weight decay of 0.0005, batch size 8, gradient-norm clipping at 1.0, and mean-squared error as the regression loss. The random seed is 2025. Training is allowed for at most 1000 epochs, with early-stopping patience of 80 epochs, a minimum improvement criterion of 1e-6, and a StepLR scheduler that halves the learning rate every 200 epochs. Leave-one-out cross-validation is performed at the complete-curve level. For the saved 99-fold leave-one-out run, the LSTM achieves a mean absolute error of 0.01373 and a coefficient of determination of 0.99570, while the GRU achieves a mean absolute error of 0.01880 and a coefficient of determination of 0.99356. These values demonstrate strong structural identifiability within the simulated dataset, but they do not constitute evidence that the physical radiative T-RNN implements LSTM or GRU dynamics.

The bidirectional software architecture is also interpreted carefully. It processes a completed temperature-indexed curve in both directions and is therefore appropriate for offline inverse identification. It is not equivalent to causal physical thermal recurrence in time. Future robustness analysis will test noisy heat flux, gap uncertainty, material-property uncertainty, and distribution shift between simulated and experimental curves. Derivative channels can enhance sensitivity to slope and curvature, but they can also amplify measurement noise.

**Table 2. Distinction between the physical radiative T-RNN and the software inverse-identification models.**

| Aspect | Physical radiative T-RNN | Software bidirectional LSTM/GRU |
|---|---|---|
| Signal carrier | Near-field radiative heat flux / temperature state | Numerical floating-point sequence |
| Recurrence | Thermal inertia, feedback, PCM history | Software hidden-state recurrence |
| State variables | $T, Q, s, f_{\mathrm{PCM}}$ | LSTM cell/hidden states or GRU hidden state |
| Couplings | $W^{\mathrm{rad}}$ constrained by geometry/materials | Trainable numerical matrices |
| Directionality | Causal physical time evolution | Bidirectional processing of a complete Q-T curve |
| Output | Thermal/logic state | Predicted filling ratio $f_g$ |
| Training | Offline physical-parameter optimization in this work | Standard software optimization |
| Claim supported | Theoretical physical recurrent architecture | Inverse identification from simulated data |

# Robustness, Physical Constraints, Scalability, and Practical Feasibility

The theoretical network isolates the radiative mechanism, but practical implementation will be affected by gap variation, PCM variability, thermal noise, parasitic conduction, nonlocal radiative coupling, many-body effects, programming resolution, and readout constraints. We denote the aggregate normalized perturbation of the physical weight matrix by $\Delta\mathbf{W}_{\mathrm{tot}}$, which collects gap-, PCM-, many-body-, cross-talk-, conduction-, and fabrication-induced contributions. Robustness can therefore be written as a perturbation of the physically realizable radiative network.

$$\mathbf{W}^{\mathrm{imp}} = \mathbf{W}^{\mathrm{rad}} + \Delta\mathbf{W}_{\mathrm{tot}} \quad (50)$$

The implementation error relative to the ideal radiative design can be quantified as $\mathcal{E}_W^{\mathrm{imp}} = \| \mathbf{W}^{\mathrm{imp}} - \mathbf{W}^{\mathrm{rad}} \|_F / (\| \mathbf{W}^{\mathrm{rad}} \|_F + \delta_{\mathrm{num}})$, while the final error relative to the target mathematical operation is $\mathcal{E}_W^{\mathrm{tar}} = \| \mathbf{W}^{\mathrm{imp}} -$

$\mathbf{W}^{\text{tar}} \|_F/(\| \mathbf{W}^{\text{tar}} \|_F + \delta_{\text{num}})$. For logic, a sufficient first-order robustness condition is $|\delta Q_i| < M_{Q,i}$. If heat-flux perturbations are characterized statistically by $\sigma_{Q,i}$, the ratio $M_{Q,i}/\sigma_{Q,i}$ measures separation from the nearest switching threshold.

Gap variation is especially important because the evanescent transmission contains $\exp(-2\kappa d)$. Small deviations $\delta d$ therefore produce $\delta Q \approx (\partial Q/\partial d)\delta d$ and $\delta W \approx (\partial W^{\text{rad}}/\partial d)\delta d$. Strong near-field gap dependence is simultaneously a source of tunability and a fabrication challenge. PCM uncertainty enters through gate temperature, phase fraction, dielectric response, and hysteresis. The first-order variance of a programmed weight depends on the sensitivities of $W^{\text{rad}}$ to these parameters and their variances. Distinguishable programmed states require their intended weight separation to exceed the corresponding uncertainty. In a dense integrated array, pairwise radiative coefficients may be modified by multiple scattering, screening, and nonlocal coupling. A full many-body calculation is therefore a future requirement for validating the pairwise baseline. Similarly, the locality of a physical convolution requires unintended coupling outside the target receptive field to remain small compared with the desired local coupling.

Lateral conduction cannot be assumed absent in a fabricated system. The complete energy balance contains both radiative and conductive terms. A useful comparison is the ratio $\Lambda = \frac{G^P_{\text{cond}}}{G^P_{\text{rad}}+\delta_{\text{num}}}$. The baseline framework intentionally isolates the radiative mechanism and conduction. System energy must include programming, gate-temperature control, thermal-state preparation, readout, and external control:

$$\begin{aligned} E_{\text{sys}} = \quad & E_{\text{write}} + E_{\text{gate}} + E_{\text{thermal}} \\ & + E_{\text{read}} + E_{\text{control}} \end{aligned} \tag{51}$$

Thus, the internal vacuum-gap information-transfer channel may be charge-carrier-free, but the complete platform may still require external electronics and energy. Likewise, the usable response time is constrained by thermal relaxation, PCM kinetics, and readout rather than by electromagnetic propagation alone. Scaling introduces further trade-offs. A fully connected N-node pairwise network contains $O(N^2)$ possible interactions, whereas a local convolution with fixed $K_x \times K_y$ receptive field scales approximately with the number of output nodes. However, the locality advantage is meaningful only if nonlocal electromagnetic cross-talk remains controlled. Increasing physical depth can increase processing capability but also accumulate thermal delay, heat-flux attenuation, parameter errors, and state-contrast loss. The maximum useful depth therefore is defined by maintaining acceptable logic margin, feature-map fidelity, and latency rather than by assuming that deeper is always better. The physical programmability of a large weight set can be analyzed through the Jacobian $J_W = \partial W^{\text{rad}}/\partial\theta$. If $\text{rank}(J_W)$ is smaller than the number of desired independent weights, complete independent programming is impossible. The condition number of $J_W$ can quantify how ill-conditioned the physical programming problem becomes. The most credible application space is specialized thermal or infrared information processing rather than universal electronic replacement. Near-field radiative coupling is attractive where non-contact transfer, spectral selectivity, and material-state-dependent coupling are useful; conventional electronics remains superior in speed, integration maturity, and general-purpose interconnection.

## Discussion

The central objective of this work is to establish a physically grounded framework in which near-field radiative heat transfer is used not only as an energy-transfer mechanism but also as an information-processing medium. The significance of the proposed architecture does not lie in the general idea that heat can perform logic operations, which has already been explored through conductive, phase-change, thermal-electric, and other approaches. Rather, the distinctive element considered here is programmable near-field electromagnetic coupling between spatially separated thermal nodes as the physical mechanism for transmitting, weighting, transforming, and retaining thermal information. The computational mapping begins from fluctuational electrodynamics rather than from neural-network equations introduced independently of heat-transfer physics. Temperature determines the thermal occupation of electromagnetic modes; the optical and phase states of the materials determine the reflection and transmission coefficients; these quantities determine the heat flux; and the local differential heat-flux response defines the physically realizable coupling. The network description therefore emerges from a physical hierarchy rather than from a direct replacement of electronic variables by thermal symbols. The work proposes and theoretically analyzes a physically constrained route for implementing neural-network-inspired operations through near-field radiation. It does not yet establish an experimental radiative deep-learning processor, autonomous thermal backpropagation, or a general-purpose replacement for electronics. This narrower positioning makes the contribution more defensible while preserving its central physical novelty.

The near-field regime matters because it provides strongly geometry-, spectrum-, material-, and state-dependent coupling. This dependence enables tunable thermal weights but simultaneously creates sensitivity to fabrication and environmental variation. The same gap dependence that makes the coupling programmable also creates a stringent tolerance problem. Practical operation therefore requires the intentional modulation of the radiative weight to remain larger than the unintentional variation produced by gap error, temperature fluctuation, material uncertainty, and cross-talk.

The physical T-CNN becomes meaningful only when an actual local convolution is demonstrated. A thermal map that appears edge enhanced is not sufficient by itself. The stronger test is a target mathematical kernel mapped into physically attainable radiative couplings, followed by direct comparison of target and physical feature maps. The physical constraints on the kernel, including signed differential encoding and approximate weight sharing, distinguish this architecture from a conventional software CNN. The physical T-RNN similarly becomes stronger when recurrence is derived from thermal balance and phase-state dynamics rather than imposed by analogy. Finite thermal capacitance creates volatile history dependence, $VO_2$ hysteresis can preserve information about the preceding heating or cooling path, and GST may offer a route to persistent phase-state storage. These mechanisms possess distinct timescales and retention semantics. Their physical diversity may ultimately be more interesting than exact imitation of software recurrent gates. The software inverse-identification study remain clearly separate. The bidirectional LSTM and GRU use simulated Q–T curves to estimate filling ratio and therefore demonstrate that near-field radiative responses encode structural information. They do not demonstrate that the physical radiative T-RNN implements LSTM or GRU equations. Treating the software model as a complementary inverse problem strengthens the paper by separating two directions of information flow: structure and state toward thermal processing, and thermal response back toward structural identification.

Conductive pathways are often easier to fabricate and may be useful for biasing, stabilization, and thermal management. Near-field radiation provides a different capability: non-contact, spectrally selective, phase-state-dependent coupling. A future practical architecture may combine the two mechanisms rather than attempt to eliminate conduction completely. [38-40] Similarly, the proposed network is not expected to outperform electronics in general switching speed or integration maturity. Its most credible opportunities arise when the input is already thermal or infrared, when local non-contact coupling is desirable, or when phase-change-dependent optical states can provide useful local programmability and memory. In such cases, thermal-domain preprocessing before full electronic readout may become valuable. The present framework establishes a bottom-up connection from fluctuational electrodynamics to thermal device physics, radiative logic, spatial convolution, and temporal recurrence. The strongest conceptual question is therefore not whether a conventional CNN or RNN can simply be copied with thermal devices, but which neural-network-like operations emerge naturally from programmable near-field heat-transfer physics and what physical constraints distinguish those operations from their software counterparts.

Future progress will proceed from carefully characterized primitives toward small, coupled networks. Experimental mapping of gate temperature, phase fraction, and gap distance to radiative weights will precede large-scale integration. A small physical kernel should then be calibrated and compared against a target convolution, and a recurrent device will be fabricated and tested using two different histories that reach the same read temperature but retain different radiative outputs. These steps would establish the core operations before any claim of scalable thermal neural processing. The broader implication is that the governing laws of heat transfer can themselves define a constrained computational space. In this space, weights are electromagnetic couplings rather than arbitrary parameters, nonlinearities arise from material optical response, memory arises from thermal and phase-state dynamics, and limitations are set by thermodynamics, fabrication, and many-body interaction. The resulting architecture is therefore best viewed as a theoretical and numerical foundation for programmable near-field radiative thermal computation and neural-network-inspired processing.

## Conclusion

We have developed a theoretical and numerical framework for near-field radiative thermal computation in which the transport, modulation, and retention of thermal information are derived from physically constrained radiative interactions. The framework begins from fluctuational electrodynamics, uses heat flux as the propagated thermal signal, defines radiative weights from differential heat-flux conductances, and introduces phase-change-controlled nonlinear modulation and tri-state heat-flux decoding. At the device level, near-field radiative thermal diodes, gate-controlled radiative modulators, and multi-terminal logic units provide directional response, programmable coupling, and logic fusion. At the network level, a physical T-CNN is formulated through finite local receptive fields and physically attainable radiative kernels, while a physical T-RNN is derived from thermal energy balance, radiative feedback, finite thermal relaxation, and PCM state evolution. The recurrent formulation distinguishes volatile thermal inertia, $VO_2$ hysteretic memory, and possible GST nonvolatile phase storage. A central outcome of the framework is the separation between physical forward computation and numerical optimization. Radiative inference may occur through the thermal network itself once the physical parameters are programmed, whereas gradient-based training is treated as an offline numerical method unless an independent in-situ thermal learning mechanism is demonstrated. A separate bidirectional LSTM/GRU inverse-identification study is likewise interpreted as a software application to simulated radiative data rather than as proof that the physical T-RNN implements software recurrent gates.

The present work remains a theoretical and simulation-based study. It does not yet demonstrate a fabricated integrated radiative processor, experimentally verified convolutional weight sharing, autonomous physical backpropagation, or universal advantages over electronic hardware. Practical realization will require quantitative treatment of nanoscale-gap tolerances, many-body electromagnetic interaction, conductive parasitics, thermal noise, PCM variability, programming and readout energy, memory retention, and scaling. Within these limitations, the framework provides a physically grounded path from near-field heat transfer to spatial and temporal information processing. Its principal value lies not in replacing electronics universally, but in exploring computation that is native to the thermal and infrared domain, where non-contact radiative coupling, spectral selectivity, and phase-state-dependent material response may provide specialized functionality. The work therefore establishes a foundation for future experimental studies of programmable near-field radiative logic, convolution, memory, and thermal-domain inference.

## Methods

The numerical framework is organized around a unified set of physical variables and a clear separation between radiative power $P$ and radiative heat flux $Q$. The symbols $x$, $u$, $h$, $u^F$, and $y$ denote, respectively, normalized temperature input, normalized physical preactivation, normalized physical heat-flux state, fused physical preactivation, and decoded tri-state logic. The generic physical programming vector is $\boldsymbol{\theta}$. The grating filling ratio is $f_g$, while the phase-change material state is represented by $f_{\mathrm{PCM}}$. The near-field radiative heat flux is evaluated with the multilayer $T_{\mathrm{quad}}$ solver used in the accompanying MATLAB implementation. At each wavelength, TE and TM reflection coefficients are obtained for the two multilayer bodies and combined into propagating- and evanescent-wave transmission factors. Writing $q = k_{\parallel}/k_0$, the propagating sector is evaluated from $q = 0$ to 0.99999 with $\Delta q = 0.001$. The evanescent sector is evaluated adaptively from $q = 1$ to $\infty$ using MATLAB's integral routine with relative tolerance $5 \times 10^{-2}$; thus, no finite $k_{\parallel}$ cutoff is imposed in the implemented solver. The spectral heat-flux integral is evaluated on a uniform angular-frequency grid with 200 intervals. The physical transistor calculations use the nominal wavelength interval 0.5–30 $\mu$m, whereas the inverse-identification dataset uses 2–80 $\mu$m. The symmetrized harmonic-oscillator energy $(\hbar\omega/2)\coth[\hbar\omega/(2k_BT)]$ is used in the code; its zero-point contribution cancels in the temperature difference, consistent with the net-flux formulation used in the theory section.

The physical transistor geometry uses a one-dimensional grating period $\Lambda = 50$ nm and a representative grating filling ratio $f_g = 0.3$. The source and drain stacks each contain a 1 $\mu$m BN layer and a 1 $\mu$m Au layer, while the gate stack uses a 0.5 $\mu$m PCM grating together with a 1 $\mu$m underlying PCM film. The vacuum gap passed to the near-field transmission solver is $d_{ij} = 50$ nm for each source–gate or gate–drain interaction in the transistor calculation. A separate 100 nm geometry variable appears in the device scripts and a 100 nm terminal-to-gate spacing was also used in the previously published multi-terminal NRTLC design [46]; these two conventions is therefore not conflated. Unless otherwise stated for a figure-specific schematic, the transistor-based coupling calculations in the present manuscript use the 50 nm single interaction gap.

The optical models are taken directly from the dielectric-function routines used by the simulations. Insulating $VO_2$ is treated anisotropically with separate ordinary and extraordinary multi-oscillator Lorentz responses. For the ordinary branch, $\varepsilon_\infty = 10$ with resonance wavenumbers $[189,270,310,340,505,600,710,10000]$ cm$^{-1}$, oscillator strengths $[0.54,13,7,0.7,3.1,4.8,0.15,1.3]$, and, fractional damping factors $[0.012,0.07,0.05,0.024,0.07,0.074,0.06,0.4]$. For the extraordinary branch, $\varepsilon_\infty = 9.7$ with resonance wavenumbers $[227.5,285,324,355,392.5,478,530,700,10000]$ cm$^{-1}$, strengths $[0.1,3.3,1.95,7.4,1.0,0.2,0.65,0.25,1.3]$, and fractional damping factors $[0.02,0.06,0.018,0.08,0.03,0.08,0.045,0.055,0.4]$. Metallic $VO_2$ is implemented with the code's Drude-like form using $\varepsilon_\infty = 9$, $\omega_p = 8000$ cm$^{-1}$, and $\omega_c = 10000$ cm$^{-1}$. Amorphous and crystalline GST are obtained from tabulated complex refractive-index data and converted through $\varepsilon = (n + i\kappa)^2$ using gridded interpolation; the dataset corresponds to the optical constants reported by Frantz et al. [58]. Au is represented by Johnson–Christy optical data below 1.9 $\mu$m with a Drude continuation at longer wavelengths [59].

Finite-width phase transitions in the static forward model are implemented with explicit hyperbolic-tangent phase fractions. For $VO_2$ over 341–346 K, the high-temperature phase fraction is $f_{VO_2}(T) = 1/2\,\{1 + \tanh[(T - 343.5)/0.5]\}$. For GST over 432–452 K, $f_{GST}(T) = 1/2\,\{1 + \tanh[(T - 442)/2]\}$. The intermediate dielectric response is generated by Maxwell–Garnett mixing between the low- and high-temperature optical states and is subsequently passed to the second-order grating EMT. These one-direction interpolation functions describe the static forward-property calculation; they are intentionally distinguished from the history-dependent heating/cooling branches introduced for the physical T-RNN memory model.

**Table 3. Material, geometrical, and numerical parameters used in the NFRHT simulations.**

| Item | Implementation | Numerical value / model | Scope |
|---|---|---|---|
| Grating model | Second-order 1-D EMT [57] | $\Lambda = 50$ nm; $f_g = 0.3$ for the representative transistor | Physical device |
| Layer thickness | Source/drain: BN + Au; gate: PCM grating + PCM film | $1 + 1$ $\mu$m; $0.5 + 1$ $\mu$m | Physical device |
| Vacuum gap | Gap passed to $T_{\text{quad}}$ | $d_{ij} = 50$ nm | Physical transistor coupling |
| Frequency sampling | Uniform angular-frequency summation | 200 intervals | Physical and inverse calculations |
| Physical spectral window | Nominal wavelength interval | 0.5–30 $\mu$m | Transistor/PCM forward calculation |
| In-plane wavevector | PW grid + adaptive EW integral | $\Delta q = 0.001$ for $0 \le q < 1$; $1 \le q < \infty$, RelTol $= 0.05$ | NFRHT solver |
| $VO_2$ transition | Tanh mixed-phase interpolation | 341–346 K; center 343.5 K; scale 0.5 K | Static forward model |
| GST transition | Tanh mixed-phase interpolation | 432–452 K; center 442 K; scale 2 K | Static forward model |
| GST optical data | Tabulated $n$ and $\kappa$ with $\varepsilon = (n + i\kappa)^2$ [58] | Tabulated range 0.35028–29.628 $\mu$m | aGST/cGST |

| Item | Implementation | Numerical value / model | Scope |
|---|---|---|---|
| Inverse dataset | Second-order EMT; sharp transition at $T_0 = 341$ K | $f_g = 0.01$–$0.99$ (100 curves); 331–351 K (500 points); $d = 50$ nm; 2–80 $\mu$m | Software inverse-identification forward data |

For multi-terminal networks, the baseline model uses pairwise radiative powers summed at each node. The net radiative power received by node $i$ is $P_i^{\text{rad}} = \sum_{j\neq i} P_{j\rightarrow i}$, and the corresponding heat flux is $Q_i^{\text{net}} = P_i^{\text{rad}}/A_i$. The complete transient node balance is $C_i \ dT_i/dt = P_i^{\text{ext}} + P_i^{\text{rad}} + P_i^{\text{cond}} - P_i^{\text{loss}}$. Pairwise calculations should be checked against energy conservation, and a many-body treatment should be introduced in future work when multiple scattering or screening becomes significant.

Physical radiative weights are obtained from the differential heat-flux conductance $G_{ij}^{Q} = \partial Q_i^{\text{net}}/\left.\partial T_j\right|_{\text{op}}$ and normalized as $W_{ij}^{\text{rad}} = G_{ij}^{Q} T_{\text{ref}}/Q_{\text{ref}}$. Target numerical weights $W_{ij}^{\text{tar}}$ are mapped to physical parameter vectors $\boldsymbol{\theta}$ by minimizing the mismatch between target and physically attainable couplings subject to $\boldsymbol{\theta} \in \Omega_{\text{phys}}$. Signed effective weights use differential positive and negative channels.

The T-CNN is implemented as a local spatial operator. For each output location, a finite receptive field is defined through a kernel $K_{u,v}^{\text{rad}} = G_{u,v}^{Q} \ T_{\text{ref}}/Q_{\text{ref}}$, stride, and boundary condition. The physical preactivation field is $u_{p,q} = \sum_{u,v} K_{u,v}^{\text{rad}} \, x_{pS_x+u,qS_y+v} + b_{p,q}^{\text{th}}$. The physical nonlinear response produces $h_{p,q}$. When layers are cascaded, heat-flux output is converted to a temperature input using either the transient energy balance or the local steady approximation $x_i^{next} = \beta_i \ h_i$ with $\beta_i = R_{\text{th},i} \ A_i \ Q_{\text{ref}}/T_{\text{ref}}$. The representative T-CNN system considered in the design study is organized as a four-layer, four-row thermal network acting on a 10 × 10 thermal-pixel field. This architecture is used as a proof-of-principle spatial-processing configuration rather than as a claim of an optimized large-scale chip layout. The T-CNN optimization is performed numerically. The baseline objective combines tri-state classification, physical margin, spatial smoothness, and radiative-weight regularization. Optional terms include target-kernel mismatch, feature-map mismatch, physical-bound penalties, and cross-talk penalties. Numerical loss weights use $\rho$ subscripts to avoid conflict with the recurrent physical retention factor $\lambda_{\text{ret}}$. Gradients are propagated through the reduced forward model and mapped to physical parameters through $\partial W^{\text{rad}}/\partial\theta$.

The T-RNN is derived from the transient energy balance and PCM state evolution. The reduced thermal-storage state obeys $\tau_i^{\text{th}} \ \ ds_i/dt = -s_i + \eta_{\text{store},i} \ \ h_i$, giving $\lambda_{\text{ret},i} = \exp\left(-\Delta t/\tau_i^{\text{th}}\right)$ and $s_{i,k+1} = \lambda_{\text{ret},i} \ \ s_{i,k} + \left(1 - \lambda_{\text{ret},i}\right)\eta_{\text{store},i} \ h_{i,k}$. Physical recurrence combines present input, returned previous output, retained thermal state, and PCM phase state. $VO_2$ is treated as a history-dependent hysteretic material, while GST is considered only as a possible nonvolatile phase-storage medium unless write/reset, drift, and endurance are explicitly modeled. The representative T-RNN schematic uses a six-module recurrent configuration with explicit feedback paths. The recurrent architecture is interpreted physically through radiative feedback, thermal relaxation, and PCM-state evolution; the six-module organization is therefore a design example rather than a direct hardware implementation of software recurrent gates.

The software inverse-identification forward database is generated separately from the physical transistor sweep. The MATLAB forward generator targets filling ratios from 0.01 to 0.99 and 500 temperatures from 331 K to 351 K, using the second-order grating EMT, a 100 nm vacuum gap, 200 spectral intervals, and a wavelength interval of 2-80 micrometres. For this dataset only, VO2 is switched sharply at 341 K between the insulating and metallic optical states. The final merged CSV used by the saved PyTorch training run contains 100 complete filling-ratio curves, corresponding to 50,000 temperature-heat-flux samples. For software preprocessing, each heat-flux curve is min-max normalized independently with a denominator offset of 1e-9. NumPy gradient is then applied successively to construct first- and second-gradient shape channels, producing a 500-by-3 input sequence for each curve. The LSTM/GRU implementation uses one bidirectional recurrent layer with 32 hidden units per direction, temporal mean pooling, dropout 0.15 outside the single recurrent layer, and a linear-GELU-dropout-linear regression head followed by a sigmoid output constraint. Optimization uses AdamW with learning rate 0.001, weight decay 0.0005, batch size 8, mean-squared error loss, gradient clipping at 1.0, seed 2025, a maximum of 1000 epochs, early-stopping patience 80 with minimum improvement 1e-6, and a StepLR decay factor of 0.5 every 200 epochs. LOOCV is performed by withholding one complete curve per fold. The numerical settings reported above are the actual implementation parameters used in the supplied simulation code. The current solver separates propagating and evanescent contributions, uses the implemented relative tolerance of 0.05 for the evanescent integration, and uses 200 uniform spectral intervals in the cited forward calculations. The nominal physical-transistor spectral window extends to 30 micrometres, whereas the available GST optical table extends to approximately 29.628 micrometres; this endpoint difference is disclosed rather than silently altering the implemented range.

## Nomenclature and Physical Meaning of Symbols

The following table consolidates the principal symbols used throughout the manuscript. It replaces separate section-by-section symbol tables so that the same physical meaning is maintained globally.

| Symbol | Physical meaning | Unit |
|---|---|---|
| $P$ | Thermal/radiative power | $W$ |
| $Q$ | Thermal/radiative heat flux | $W\ m^{-2}$ |
| $T$ | Absolute temperature | K |
| $\omega$ | Angular frequency | rad $s^{-1}$ |
| $\Theta(\omega, T)$ | Mean thermal energy of a bosonic oscillator excluding zero-point term | J |
| $\hbar$ | Reduced Planck constant | J $s$ |
| $k_B$ | Boltzmann constant | J $K^{-1}$ |
| $k_\parallel$ | Wavevector component parallel to interfaces | $m^{-1}$ |
| $k_0$ | Vacuum wavenumber $\omega/c$ | $m^{-1}$ |
| $\kappa$ | Evanescent decay constant | $m^{-1}$ |
| $\xi^p$ | Mode-resolved electromagnetic energy transmission coefficient | dimensionless |
| $\Phi(\omega)$ | Spectral transmission function per unit area after in-plane-wavevector integration | $m^{-2}$ |
| $r^p$ | Polarization-resolved reflection coefficient | dimensionless |
| $d_{ij}$ | Vacuum gap between nodes $i$ and $j$ | $m$ |
| $A_i$ | Reference interaction area of node $i$ | $m^2$ |
| $C_i$ | Effective thermal capacitance | J $K^{-1}$ |
| $R_{\text{th},i}$ | Effective thermal resistance | K $W^{-1}$ |
| $G_{ij}^{Q}$ | Differential heat-flux conductance $\partial Q_i/\partial T_j$ | $W\ m^{-2}\ K^{-1}$ |
| $G_{ij}^{P}$ | Differential power conductance when a power basis is used | $W\ K^{-1}$ |
| $Q_{\text{ref}}$ | Heat-flux normalization scale | $W\ m^{-2}$ |
| $T_{\text{ref}}$ | Temperature normalization scale | K |
| $x_i$ | Normalized temperature-coded input | dimensionless |
| $u_i$ | Normalized physical preactivation / weighted radiative input | dimensionless |
| $h_i$ | Normalized continuous physical heat- | dimensionless |

| | flux state | |
|---|---|---|
| $u_i^F$ | Fused physical preactivation formed from branch outputs | dimensionless |
| $y_i$ | Decoded physical tri-state logic | −1, 0, +1 |
| $Q_{\mathrm{th}}^{\mathrm{L}}$, $Q_{\mathrm{th}}^{\mathrm{U}}$ | Lower and upper physical heat-flux thresholds | $W\ m^{-2}$ |
| $W_{ij}^{\mathrm{rad}}$ | Physically realizable radiative weight | dimensionless |
| $W_{ij}^{\mathrm{tar}}$ | Target numerical weight | dimensionless |
| $W_{ij}^{\mathrm{pos}}$, $W_{ij}^{\mathrm{neg}}$ | Positive differential physical weight branches | dimensionless |
| $\theta$ | Physical programming/design parameter vector | mixed |
| $\Omega_{\mathrm{phys}}$ | Physically admissible parameter domain | — |
| $f_{\mathrm{PCM}}$ | PCM phase fraction | dimensionless |
| $f_g$ | Geometrical grating filling ratio | dimensionless |
| $t_{\mathrm{PCM}}$ | PCM thickness | $m$ |
| $T_S, T_G, T_D$ | Source, gate, and drain temperatures | K |
| $\chi_G$ | Drain heat-flux sensitivity to gate temperature | $W\ m^{-2}\ K^{-1}$ |
| $\mathcal{R}_D$ | Thermal rectification coefficient | dimensionless |
| $K_{uv}^{\mathrm{rad}}$ | Physical radiative convolution-kernel coefficient | dimensionless |
| $\mathbf{K}^{\mathrm{tar}}$ | Target mathematical convolution kernel | dimensionless |
| $S_x, S_y$ | Spatial convolution strides | nodes |
| $\mathcal{N}_{p,q}$ | Local receptive-field neighborhood of output position (p,q) | — |
| $\mathcal{E}_K$ | Relative physical kernel error | dimensionless |
| $\mathcal{E}_{\mathrm{map}}$ | Relative feature-map error | dimensionless |
| $\tau_i^{\mathrm{th}}$ | Physical thermal relaxation time | $s$ |
| $\tau_i^{\mathrm{PCM}}$ | Characteristic PCM evolution time | $s$ |
| $s_i$ | Retained normalized thermal state | dimensionless |
| $\eta_{\mathrm{store},i}$ | Heat-flux-to-storage coupling factor | dimensionless |
| $\lambda_{\mathrm{ret},i}$ | Discrete physical retention factor $\exp(-\Delta t/\tau_i^{\mathrm{th}})$ | dimensionless |
| $\mathcal{H}_i^{\mathrm{hist}}$ | Thermal-history state descriptor | — |
| $M_H$ | Normalized radiative separation between two thermal histories | dimensionless |
| $\mathcal{M}_i^{\mathrm{th}}(\Delta t)$ | Normalized thermal retention function | dimensionless |
| $E_{\mathrm{write}}, E_{\mathrm{reset}}$ | Write and reset energies | J |
| $\rho_r$ | Generic numerical objective/regularization coefficient | dimensionless |

| $\alpha_{\text{opt}}$ | Numerical optimization step size | parameter-scaled |
|---|---|---|
| $\mathcal{L}$ | Generic numerical loss/objective | dimensionless after normalization |
| $f_g^{\text{pred}}$ | Software-predicted geometrical filling ratio | dimensionless |
| MAE | Mean absolute error of filling-ratio prediction | dimensionless |
| $R^2$ | Coefficient of determination | dimensionless |
| $\sigma_Q$ | Heat-flux uncertainty | $W\ m^{-2}$ |
| $\chi_{\text{XT}}$ | Radiative cross-talk ratio | dimensionless |
| $\mathcal{E}_{\text{bal}}$ | Normalized energy-balance error | dimensionless |
| $\mathcal{F}_i^{\text{act}}$ | Physical nonlinear activation/response function | — |
| $\gamma_r$ | Fusion coefficient for branch r | dimensionless |
| $b_i^{\text{th}}$ | Normalized thermal bias in a preactivation | dimensionless |
| $\eta_D$ | Reduced reverse-response/leakage factor of diode model | dimensionless |
| $\alpha_T$ | Differential thermal amplification coefficient when evaluated | dimensionless |
| $\pi_{i,c}$ | Numerical SoftMax probability for class c; not physical power | dimensionless |
| $\ell_{i,c}$ | Numerical class logit | dimensionless |
| $\delta_{\text{num}}$ | Small dimensionless numerical stabilization constant | dimensionless |
| $\beta_i$ | Heat-flux-to-temperature interlayer transduction factor | dimensionless |
| $L_i^{\text{lat}}$ | PCM latent heat per unit mass in the extended energy balance | J $\text{kg}^{-1}$ |
| $\Delta\mathbf{W}_{\text{tot}}$ | Aggregate normalized perturbation of the physical radiative weight matrix | dimensionless |
| $\Lambda$ | Conductive-to-radiative power-conductance ratio | dimensionless |
| $\mathbf{J}_W$ | Jacobian of physical weights with respect to programming parameters | parameter-dependent |
| $\mathcal{D}_Q$ | Physical tri-state heat-flux decoder | — |

## Data Availability

The data supporting the findings of this theoretical and numerical study are available from the corresponding author upon reasonable request.

## Code Availability

The code used to generate and analyze the numerical data is available from the corresponding author upon reasonable request.

## Acknowledgements

This work was financially supported by the National Science Foundation (Grant No. CBET-1941743).

## Author Contributions

H.Z. designed the project and developed the theoretical model. Y.Z. supervised the project. H.Z., Y.Z., and M.A. discussed the results. H.Z. and Y.Z. wrote the manuscript. All authors revised the manuscript.

## Competing Interests

The authors declare no competing interests.